\documentclass[10pt,preprint]{aastex}

\usepackage{graphicx}
\usepackage{epsfig}
\usepackage{natbib}

\begin{document}
\title{SEISMIC CONSTRAINTS ON INTERIOR SOLAR CONVECTION}
\author{Shravan M. Hanasoge\altaffilmark{1,2}, Thomas L. Duvall, Jr.\altaffilmark{3}, \& Marc L. DeRosa\altaffilmark{4}}
\altaffiltext{1}{Max-Planck-Institut f\"{u}r Sonnensystemforschung, Max Planck Stra{$\beta$}e 2, 37191 Kaltenburg-Lindau, Germany}
\altaffiltext{2}{Department of Geosciences, Princeton University, Princeton, NJ 08544, USA}
\altaffiltext{3}{Solar Physics Laboratory, NASA/Goddard Space Flight Center, Greenbelt, MD 20771}
\altaffiltext{4}{Lockheed Martin Solar \& Astrophysics Laboratory, Palo Alto, CA 94304, USA}

\email{hanasoge@mps.mpg.de}
\begin{abstract}
We constrain the velocity spectral distribution of global-scale solar convective cells at depth using techniques of local helioseismology.
We calibrate the sensitivity of helioseismic waves to large-scale convective cells in the interior by analyzing 
simulations of waves propagating through a velocity snapshot of global solar convection via methods of time-distance helioseismology. 
Applying identical analysis techniques to observations of the Sun, we are able to bound from above the magnitudes of solar convective cells as a function of spatial convective scale.
We find that convection at a depth of $r/R_\odot = 0.95$ with spatial extent $\ell <20$, where $\ell$ is the spherical harmonic degree, comprise weak flow 
systems, on the order of 15 m/s or less. Convective features deeper than $r/R_\odot = 0.95$ are more difficult to image due to the rapidly decreasing sensitivity of helioseismic waves.
\end{abstract}

\section{Introduction \& Methodology}
Constraining the length-scales and velocities of solar interior convection is an important step towards testing and improving models of astrophysical convection. In this regard, 
two observational efforts have received much attention: the detection of the elusive Ôgiant cellsÕ \citep[e.g.][]{brown, chiang, wilson, duvall2, featherstone} 
and a study of supergranulation and near-surface convection \citep[e.g.][]{zhao, woodard}. Despite these numerous investigations, neither have giant cells been conclusively observed, 
nor have the flow systems beneath supergranules been convincingly imaged. In particular, the inversions for supergranular flow are highly susceptible to the systematical issue of cross-talk, a
situation where different velocity components possess similar signatures in the observed time shifts. Questions relating to bounds on the degree of detectability 
of large-scale convection are also not new \citep[e.g.][]{ball, swiss, hanasoge07a}. 

The Anelastic Spherical Harmonic (ASH) simulations of solar convection in a spherical shell have provided us with a computational picture of the dynamical appearance, evolution, and 
disappearance of giant convective cells \citep{miesch}. These simulations encompass a large fraction of the solar convection zone, spanning 0.7 $R_\odot$ to 0.98 $R_\odot$.  Because of 
the large disparity of time scales between the upper and lower regions of the convection zone, \citet{miesch} invoke the anelastic approximation \citep{gough} to render the problem computationally tractable
and to limit the computational cost. The anelastic formulation of the Navier-Stokes equations disallows acoustic waves; since it can only capture the dynamics that are subsonic, the domain of computation 
is truncated at 0.98 $R_\odot$ because the 
near-sonic and supersonic fluid motion contained within the near-photospheric layers of the Sun would not be realistically captured.  In spite of the use of physically unrealistic boundary conditions (i.e. impenetrable 
walls at the upper and lower radial boundaries), and parameters (e.g. Prandtl, Rayleigh, and Reynolds numbers) that differ markedly from those describing the Sun, these models provide the best 
indications at present of what transpires in the solar convective interior.



\begin{figure}[!ht]
\begin{centering}
\epsscale{1.}
\vspace{-1.5cm}
\plotone{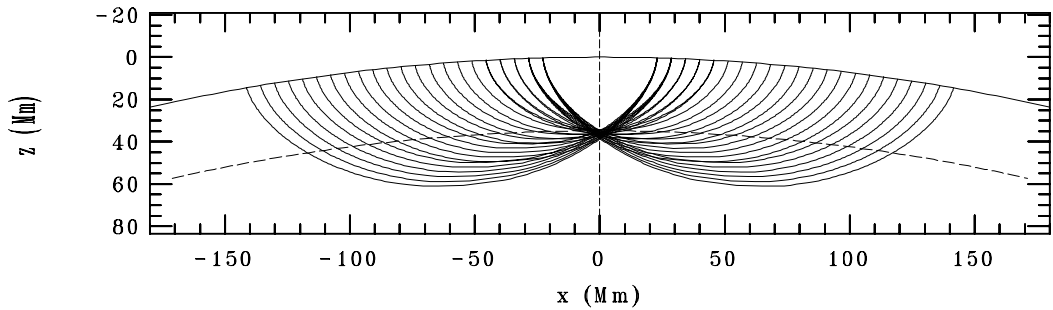}
\vspace{-1.5cm}
\caption{A pictorial representation of deep-focusing time-distance helioseismology. Numerous waves, denoted here by rays, that intersect at $r/R_\odot = 0.95$ are utilized in order to image flows 
at that depth (shown by the horizontal curved dashed line) and that horizontal location. The signal associated with the waves is measured at the solar photosphere (depicted by the horizontal curved solid line) 
\label{geometry}}
\end{centering}
\end{figure}

Acoustic waves are our primary source of information about the optically thick solar interior. Therefore, in order to begin the task of understanding the influence of convection 
on the waves, we insonify the ASH simulations and characterize the impact of the convective velocities on the wave-field statistics. In order to accomplish this, we employ the forward 
modeling techniques of Hanasoge et al. (2006), who have devised a computational means of studying the impact of thermal and flow perturbations on the acoustic wave field. We place 
a velocity snapshot of the convection zone from an ASH calculation in a solar-like stratified background and simulate wave propagation through this complex medium
\citep[cf. Figures 1 and 2 of][]{miesch08}. 
We apply the method of time-distance helioseismology (Duvall et al. 1993), which primarily utilizes temporal correlations of velocity signals at spatially disparate points in order to construct 
the statistics of the wave field. In particular, we use the technique of deep focusing \citep{jensen01}, which relies on an elegant choice of observation points, leveraged in a manner so as to 
optimize the imaging capability, as illustrated in Figure~\ref{geometry}. This method allows us to image the three components of the background velocity field, i.e., the latitudinal, longitudinal, and radial velocities.

\section{Convection Snapshot \& Wave Calculation}

\begin{figure}[!ht]
\begin{centering}
\epsscale{1.}
\plotone{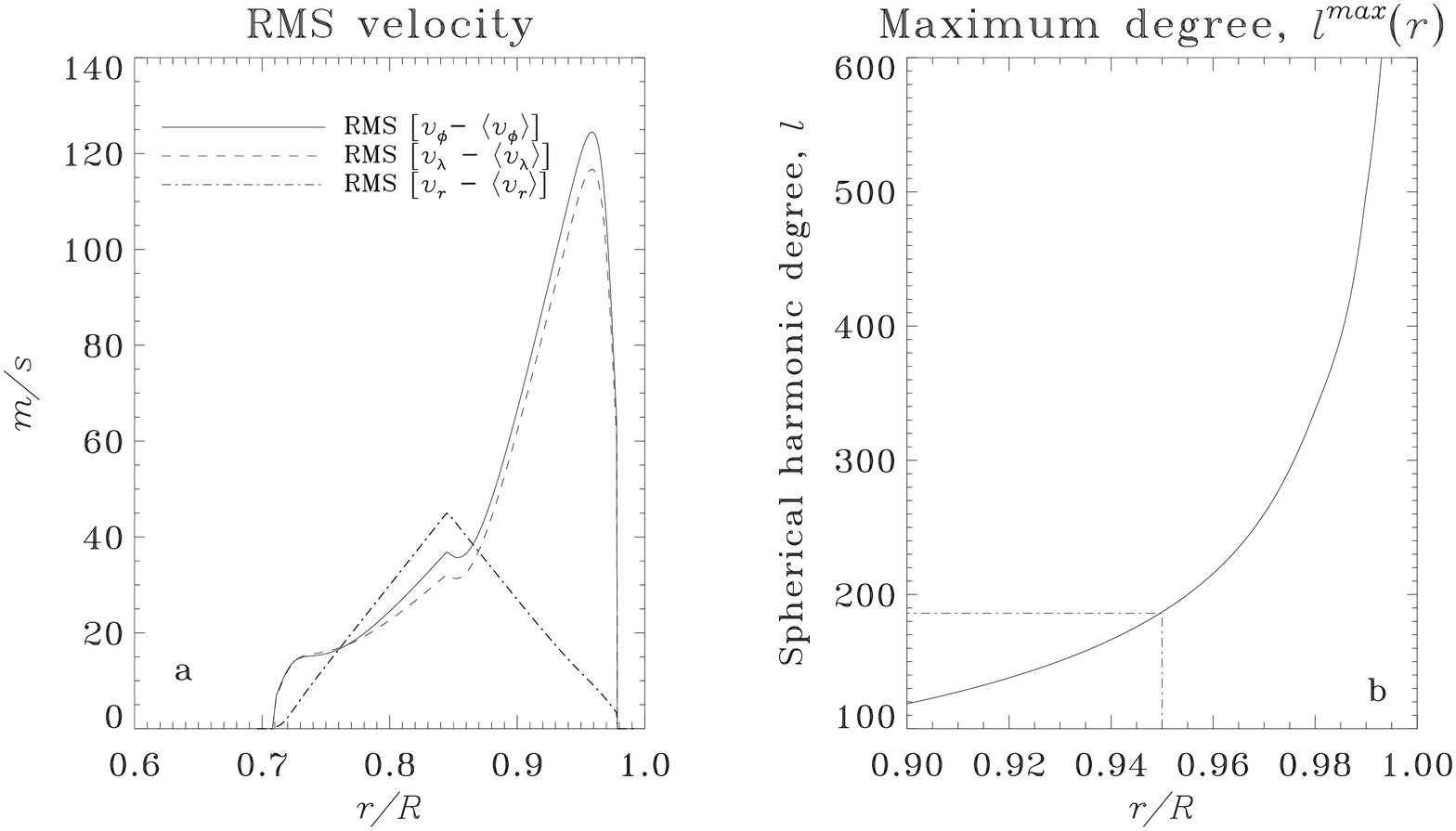}
\caption{The velocity magnitude distribution of the three interpolated velocity components derived from the ASH simulation as a function of depth (panel a). 
Longitudinal averages of the components ($\langle v_r\rangle$, $\langle v_\lambda\rangle$, $\langle v_\phi\rangle$) have been subtracted from the velocities.
 The curve $\ell^{max}(r)$ in panel (b) is the maximum wavenumber of 3.5 mHz waves that penetrate to that depth; because the imaging resolution is on the order of a wavelength, 
we are only able to image features that obey $\ell < \ell^{max}(r)$; we also have $\ell^{max}(r) \sim 185$ at $r/R_\odot = 0.95$.
\label{powermag}}
\end{centering}
\end{figure}

For purposes of discussion, we adopt a spherical co-ordinate system, using symbols $r, \lambda$, and $\phi$ to denote radius, latitude, and longitude respectively. Slices at three radii 
corresponding to $r = 0.714R_\odot, 0.85R_\odot, 0.979R_\odot$ are extracted from an ASH simulation of solar spherical convection \citep{miesch08}. Each slice consists of 
$2048\times 1024$ longitudes and latitudes respectively, with the highest spherical harmonic degree being $\ell_{\rm ASH} = 683$. Note that the number of latitudes and the maximum spherical 
harmonic degree are chosen to satisfy the relation $n_{lat} \ge 3\ell_{\rm ASH}/2$ in order that the simulation be stable against aliasing instabilities.  We linearly interpolate between the three slices
to obtain a smooth velocity field as a function of radius. The depth variation of the RMS variations in the velocity components is plotted in
Figure~\ref{powermag}(a). In panel (b), we plot the maximum wavenumber of 3.5 mHz waves that penetrate a given depth (i.e., the inner turning points of the waves); this is roughly 
estimated as $\ell^{max} = \omega r/c$, where $\omega$ is the angular frequency ($=2\pi\times0.0035$) of the waves, and $c = c(r)$ is the sound speed. 
Because the imaging resolution is on the order of a wavelength \citep[e.g.][]{gizon04}, we are unable to detect convective features of scale $\ell>\ell^{max}$ at that given depth. In the discussions 
that follow, we study the properties of convection at a depth $r/R_\odot = 0.95$, this layer being 
moderately distant from the upper boundary of the ASH simulations but not so deep that helioseismic analyses become noisy. From the curve in Figure~\ref{powermag}(b), we determine that 
$\ell^{max}(0.95R_\odot) \sim 185$, implying that convective features with $\ell > 185$ are not detectable at this depth. Thus, to moderate the computational cost, we filter the velocities in the 
spherical-harmonic  domain so as to only retain wavenumbers $\ell<256$ and then resample the data on to a grid of $768\times384$ longitudes and latitudes.


Having thus generated a 3-D cube of the convective velocities, we place it in a solar-like stratified spherical shell that extends from $r/R_\odot = 0.24$ to $r/R_\odot = 1.001$, 
and propagate waves through this medium according to the numerical recipe described in \citet{hanasoge1}. We solve the 3D linearized unsteady Euler equations describing wave propagation 
in spherical geometry in a temporally static background medium (in this case, the stratification and the convective velocities are constant in time). The waves are generated by deterministically 
forcing the radial momentum equation at a depth of approximately 100 km below the photosphere. Subsequently, a time series of the oscillation velocities are extracted at a height of 200 km above the 
photosphere and used as artificial data for helioseismic investigations. We also apply the technique of noise subtraction \citep{hanasoge07b} in order to obtain gains in the signal-to-noise ratio 
(SNR); two simulations are performed, one being a quiet calculation and the other with the convective velocity snapshot, both with identical source functions. The wave statistics of the former are 
subtracted from the latter, and the high degree of correlation between the two data sets results in a gain in the SNR and also a more direct view of the scattering process. Note that this is a luxury 
limited entirely to theoretical calculations and no observational analog exists (as yet).

\section{Time-Distance Analyses \& Caveats}
Time-distance measurements have primarily been made using cross correlations connecting points with a common surface focus \citep{Duvall1996}. This is most appropriate for studying 
near-surface phenomena, although, as \citet{giles00} has shown, it is possible to use that type of geometry to study very deep axisymmetric perturbations. A second type of geometry 
using cross correlations from pairs of points at opposite sides of a circle \citep{duvall03} would seem to be more appropriate for focusing on deep phenomena, although different distances 
then have common points at different depths. In a slightly different technique, adopted by \citet{rajaguru08}, averages of signals over a quadrant of a circle are cross correlated with the signals 
at the opposite quadrant. A possibly superior approach is to cross correlate a number of pairs of points whose connecting ray paths all converge at a subsurface focus \citep[][]{duvall94}. 
This method is computationally expensive and although it would appear to be ideal, it was shown by \citet{jensen01} that the Born sensitivity kernel corresponding to this 
geometry actually has a hollow sphere about the ÔfocusÕ point. This is consistent with the banana-doughnut nature of the two-point kernel, with zero sensitivity along the classical ray path \citep{dahlen99,birch00}.

In the present study, the latter form of deep focusing is used. At each point on the map, the input datacube is interpolated onto a spherical polar coordinate grid with the surface projection of the focus point placed 
at the north pole. The grid is computed in co-latitude from the pole (or focus point) to the largest angular distance needed for the computation. The resolution in co-latitude is the same as for 
the input data. The grid contains an equal number of longitudinal points at each latitude, making it easy to compute cross correlations between pairs of points on opposite sides of the focus 
point at different colatitudes. A ray calculation determines which pairs of points are to be used in order to focus on the desired subsurface location. The extent in co-latitude is restricted by requiring
that the rays lie within 45$^\circ$ of the horizontal tangent plane at the focus depth. In effect, this limits the distance of the antennae from the focal point.

\begin{figure}[!ht]
\begin{centering}
\epsscale{1.}
\plotone{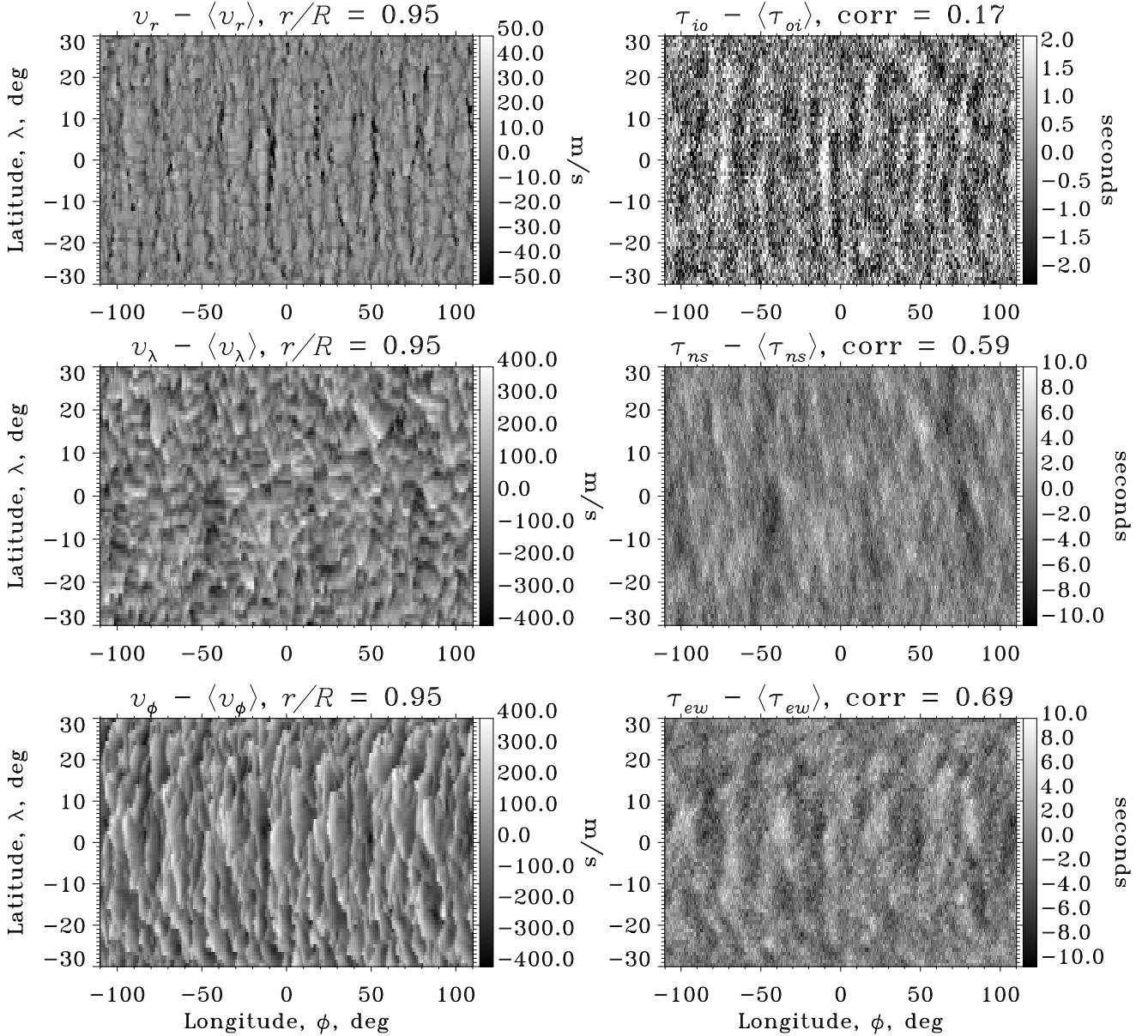}
\caption{Comparison of flows at a depth of $r/R_\odot = 0.95$ and deep-focus travel time differences (configuration shown in Figure~\ref{geometry}). The left column shows the flows from the 
simulation: from top to bottom, the radial ($v_r$), north-south ($v_\lambda$) and east-west ($v_\phi$) components at $r/R_\odot$ = 0.95. Travel time differences corresponding to these three 
components are shown on the right-hand column: from top to bottom, the difference between the ingoing and outgoing travel times ($\tau_{io}$), the north-south travel-time asymmetry ($\tau_{ns}$), 
and the east-west travel-time asymmetry ($\tau_{ew}$). The correlation coefficients between the flows and travel time maps is 0.17 for the radial, 0.59 for the north-south, and 0.69 for the 
east-west cases. The spatial cross correlations between the travel-time and velocity maps have a full-width at half-maximum of 9 deg for both the north-south and east-west cases.
\label{traveltimes}}
\end{centering}
\end{figure}

The three travel-time maps obtained subsequently can be interpreted to relate to the flows in the following manner: divergence time shift maps ($\tau_{io}$) correspond to radial flows, 
north-south time-shift maps ($\tau_{ns}$) correspond to latitudinal velocities ($v_\lambda$) and the east-west time-shift maps ($\tau_{ew}$) to longitudinal velocities ($v_\phi$). As is seen in 
Figure~\ref{traveltimes}, there is a significant resemblance between the convective velocities at depth and the associated deep-focusing time shifts. In particular, the correlation between the east-west time 
shifts and the longitudinal convective velocities is at the level of 70\%; latitudinal velocities are also strongly correlated with the north-south time shifts, at the level of 60\%. The radial flows, 
significantly weaker than the other components, do not register quite so well in the travel times.

\begin{figure}[!ht]
\begin{centering}
\epsscale{1.}
\plotone{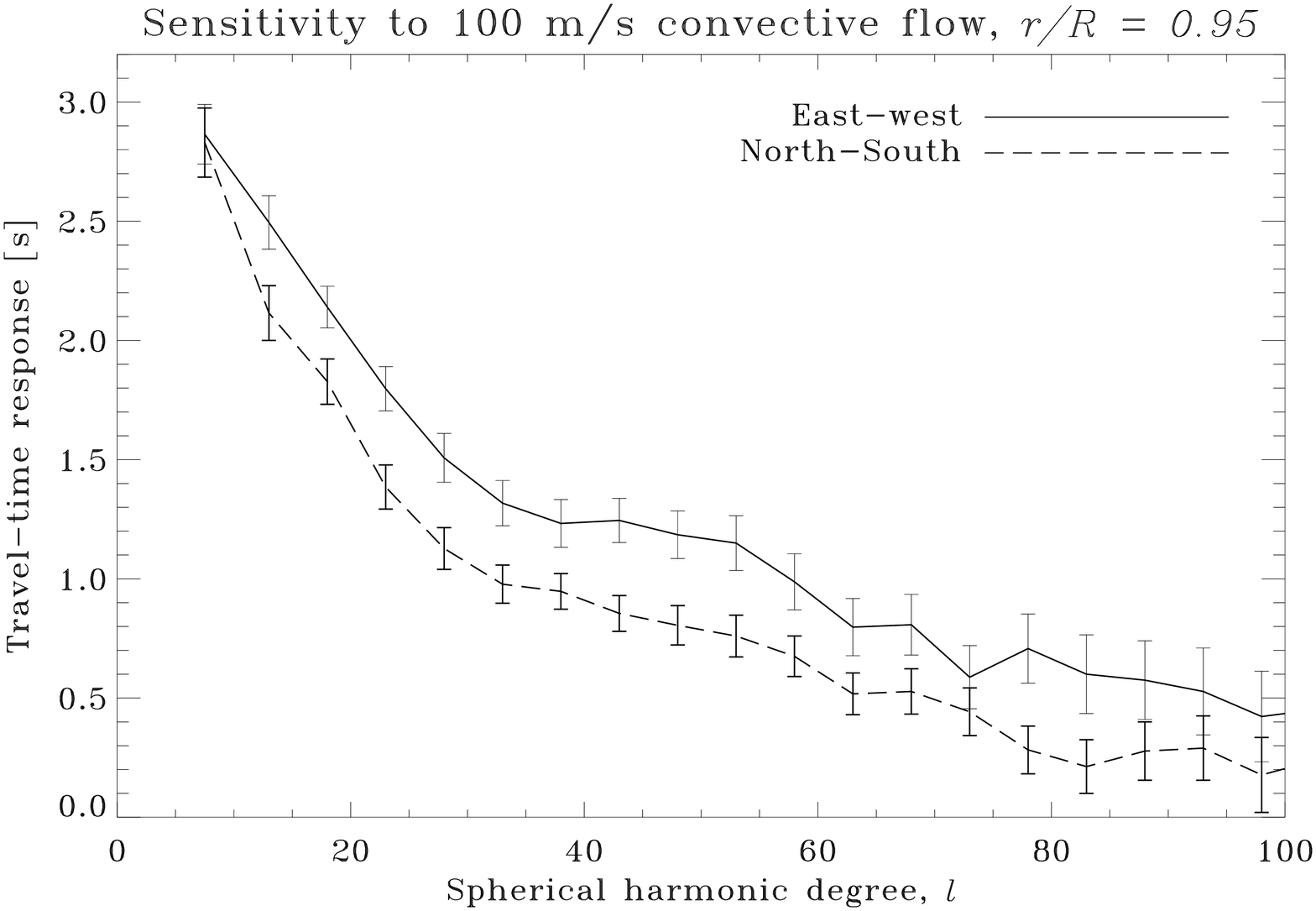}
\caption{A measure of the imaging resolution of helioseismic waves. For example, a convective cell with horizontal velocity amplitude 100 m/s and dominant spatial power in $\ell=1$ will elicit approximately a 3 second shift 
in the travel times as measured by this deep-focusing technique. Similar interpretations apply to convective features at higher $\ell$. Finer-scale features at this depth, i.e. those characterized by $\ell > 50$, 
register much more weakly in the travel times. Note that these curves suggest an imaging resolution of $\ell \sim 30$ (defined as the half-width), far smaller than the largest wavenumber  
($\ell^{max} \sim 185$) that propagates at this depth ($r/R_\odot = 0.95$).
\label{sensitivity}}
\end{centering}
\end{figure}

We now calibrate the shifts in travel times directly to the convective velocities at the focus depth ($r/R_\odot =0.95$) of the waves. We transform the velocities and time shifts into spherical
harmonic space and perform a linear fit between the real and imaginary harmonic coefficients of the velocities and time shifts as a function of the bandpass (or spherical harmonic degree, $\ell$).
We graph this sensitivity in Figure~\ref{sensitivity}.
This calibration allows us to determine the induced time shift by a convective-cell-like feature at depth; thus the observationally derived time shifts may be ``inverted" in order to determine
the convective velocity. A notable aspect of Figure~\ref{sensitivity} is that the imaging resolution (defined as the half-width) as suggested by the curves is $\ell \sim 30$, far smaller than 
$\ell^{max} \sim 185$ at this depth ($r/R_\odot = 0.95$).

Caution must be exercised when interpreting this calibration process, primarily because the convection model we employ is greatly simplified, and is likely not very comparable to the 
complex solar medium. In particular, we must keep in mind that: (1) the sensitivity model of the waves has not been carefully investigated; thus, despite using this intricate focusing geometry, 
we possibly still retain considerable sensitivity to the near-surface layers \citep[e.g. Figure 4.13 of][]{birch_thesis}, (2) we do not model the action of the near-surface convection on the waves, thereby not incorporating a significant source 
of wave scattering, (3) we linearly interpolate the convective velocity data in depth between the three available slices, thereby further reducing the complexity of the interior convection 
model, and (4) the solid-wall upper boundary condition applied in the ASH convection simulation leads to unrealistic convective velocity profiles, especially around $r/R_\odot = 0.979$.

\section{Observations \& Conclusions}
\begin{figure}[!ht]
\begin{centering}
\epsscale{1.}
\plotone{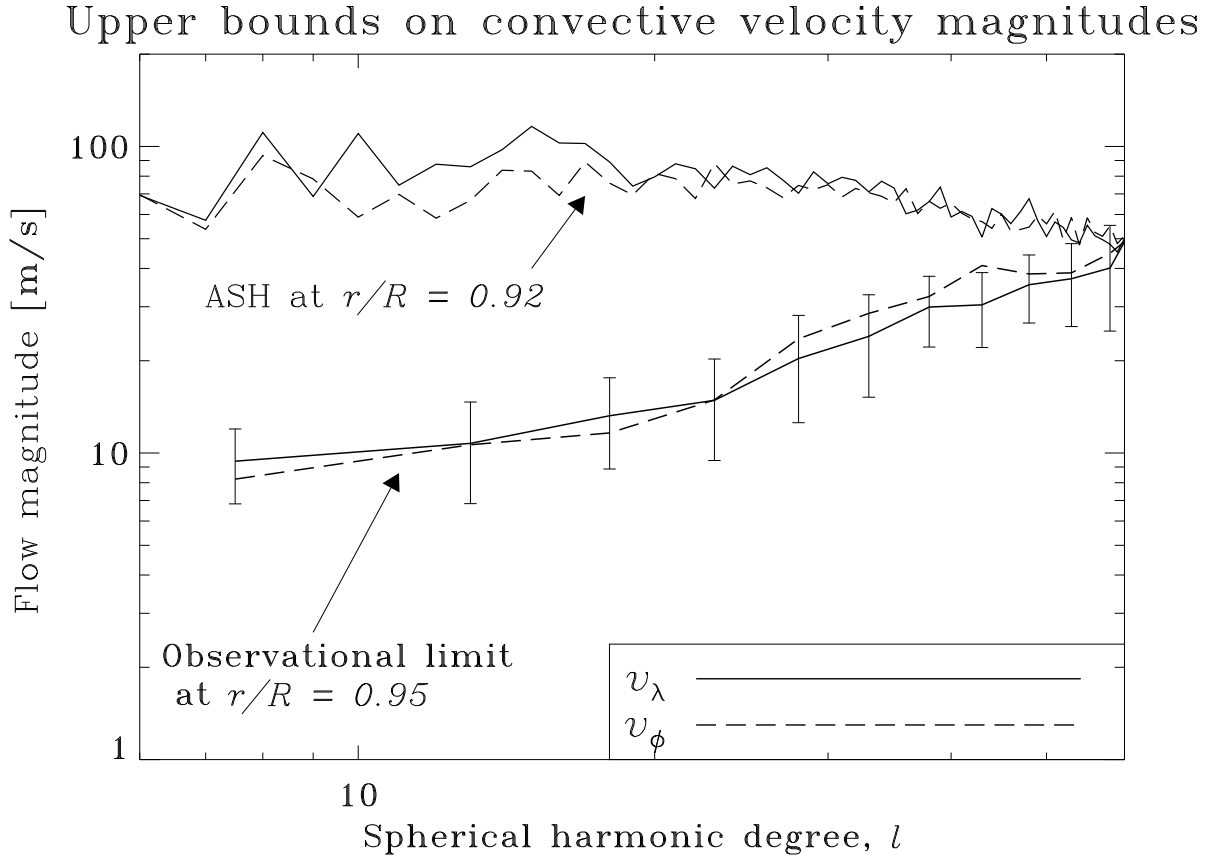}
\caption{Upper bounds on the velocity magnitude spectrum of interior convection (with differential rotation and meridional circulation removed). 
Error bars are identical for both components but for greater clarity, are only shown for the longitudinal velocity.
The average level of the travel-time noise 
in single day MDI medium-$\ell$ quiet Sun data is converted to velocity using the calibration curve in Figure~\ref{sensitivity}. Based on the non-detection of convection, 
 these form upper bounds on the velocities of convective cells. Giant cells with $\ell < 20$ at $r/R_\odot = 0.95$, if they exist, must therefore
be of velocity magnitudes on the order of 15 m/s or less, making them very difficult to detect. Also plotted are
spectra obtained from a 62 day average of the longitudinal and latitudinal convective velocities at $r/R_\odot = 0.92$ \citep[Figures 13b and c of][]{miesch08}. Note that the 
velocity magnitude increases as a monotonic function of $r/R_\odot$, implying that the ASH velocity magnitudes are even higher at $r/R_\odot = 0.95$.  Note that the seeming 
asymmetry in the error bars has to do with using a log scale. 
\label{upperbounds}}
\end{centering}
\end{figure}

Deep-focusing analyses of individual days of quiet Sun observations of MDI medium-$\ell$ data \citep{scherrer} show no indication of the presence of larger-scale convective features. 
The time shifts are dominated by noise. Furthermore, the correlation coefficient between time-shift maps of one day and the next (corrected for rotation) is too small to be regarded statistically significant.
Similar to the work of  \citet{braun08}, we utilize this non-detection to constrain the magnitudes of the convective velocities from above. We first obtain estimates of the mean level of noise 
present in the travel times as a function of spherical harmonic degree. MDI medium-$\ell$ data from the two-year quiet period October 2007 - October 2009
are analyzed and twenty-eight synoptic charts of the travel-times corresponding to as many Carrington rotation periods are constructed. Over a rotation period, a given heliocentric longitude is visible
for seven days; thus the expectation noise level associated with the chart is $\sqrt{7}$ times the value derived from one day's analysis. We multiply the chart by this number in order to restore the noise to the one-day level. 
The spherical harmonic spectra of these charts are then averaged and a mean distribution of noise as a function of spherical harmonic degree is recovered. Lastly, these expectation values for the noise are divided by the 
calibration function of Figure~\ref{sensitivity} to obtain an upper-bound estimate 
for the convective velocities (Figure~\ref{upperbounds}). 

In other words, the noise-dominated time shifts are converted to velocities, with the implication that the magnitude of the 
interior solar convective velocity spectrum can be no greater than suggested by the curve in Figure~\ref{upperbounds}, failing which large-scale convective cells become detectable 
within this observational window. As seen in the figure, the constraints 
place tight restrictions on the convective velocities, especially at low spherical harmonic degree. Giant cells at depth, if they exist, have small velocity magnitudes, on the order of 15 m/s or less, underscoring
the remarkable challenge in actually detecting them. Also plotted is the longitudinal convective velocity spectrum obtained from a 62 day average of the ASH simulations; these velocities are seen to be of 
much greater magnitude than suggested by the observations. The 1-$\sigma$ error bars in Figure~\ref{upperbounds} are estimated by propagating the errors on the sensitivities (Figure~\ref{sensitivity}) and the
variance in the one-day noise level. At each spherical harmonic degree $\ell$, we average the noise associated with $2\ell + 1$ modes (i.e. $|m| \le \ell$). Because very few modes are averaged at low $\ell$, the 
primary source of uncertainty in estimating the convective velocities is the (relatively) high degree of variance in the one-day noise level. At higher spherical harmonic degrees, the sensitivity starts to fall and the 
associated uncertainties dominate.

\acknowledgements
The simulations presented here were performed on the Schirra supercomputer at NASA Ames. S.M.H. would like to extend warm regards towards P. Scherrer for supporting a visit to Stanford over which some of
the work was accomplished. We would also like to thank M. Miesch for sending us the ASH data used in Figure~\ref{upperbounds} and J. Toomre for helpful comments.
S.M.H. acknowledges funding from the German Aerospace Center (DLR) through grant FK Z 50 OL 0801 ``German Data Center for SDO" and the Princeton University Department
of Geosciences. T.D. appreciates funding from NASA and the continued hospitality of the Stanford Solar Group.

\bibliographystyle{apj}

\begin{thebibliography}{27}
\expandafter\ifx\csname natexlab\endcsname\relax\def\natexlab#1{#1}\fi

\bibitem[{{Beck} {et~al.}(1998){Beck}, {Duvall}, \& {Scherrer}}]{duvall2}
{Beck}, J.~G., {Duvall}, Jr., T.~L., \& {Scherrer}, P.~H. 1998, \nat, 394, 653

\bibitem[{{Birch}(2002)}]{birch_thesis}
{Birch}, A.~C. 2002, PhD thesis, Stanford University, USA

\bibitem[{{Birch} \& {Kosovichev}(2000)}]{birch00}
{Birch}, A.~C., \& {Kosovichev}, A.~G. 2000, \solphys, 192, 193

\bibitem[{{Braun} \& {Birch}(2008)}]{braun08}
{Braun}, D.~C., \& {Birch}, A.~C. 2008, \apjl, 689, L161

\bibitem[{{Brown} \& {Gilman}(1984)}]{brown}
{Brown}, T.~M., \& {Gilman}, P.~A. 1984, \apj, 286, 804

\bibitem[{{Chiang} {et~al.}(1987){Chiang}, {Petro}, \& {Foukal}}]{chiang}
{Chiang}, W.-H., {Petro}, L.~D., \& {Foukal}, P.~V. 1987, \solphys, 110, 129

\bibitem[{{Duvall} {et~al.}(2001){Duvall}, {Jensen}, {Kosovichev}, \&
  {Birch}}]{jensen01}
{Duvall}, T.~L., {Jensen}, J.~M., {Kosovichev}, A.~G., \& {Birch}, A.~C. 2001,
  AGU Spring Meeting Abstracts, 22

\bibitem[{{Duvall}(1995)}]{duvall94}
{Duvall}, Jr., T.~L. 1995, in Astronomical Society of the Pacific Conference
  Series, Vol.~76, GONG 1994. Helio- and Astro-Seismology from the Earth and
  Space, ed. R.~K. {Ulrich}, E.~J. {Rhodes}, Jr., \& W.~{Dappen}, 465--+

\bibitem[{{Duvall}(2003)}]{duvall03}
{Duvall}, Jr., T.~L. 2003, in ESA Special Publication, Vol. 517, GONG+ 2002.
  Local and Global Helioseismology: the Present and Future, ed.
  H.~{Sawaya-Lacoste}, 259--262

\bibitem[{{Duvall} {et~al.}(1996){Duvall}, {D'Silva}, {Jefferies}, {Harvey}, \&
  {Schou}}]{Duvall1996}
{Duvall}, T.~L.~J., {D'Silva}, S., {Jefferies}, S.~M., {Harvey}, J.~W., \&
  {Schou}, J. 1996, \nat, 379, 235

\bibitem[{{Featherstone} {et~al.}(2006){Featherstone}, {Haber}, {Hindman}, \&
  {Toomre}}]{featherstone}
{Featherstone}, N.~A., {Haber}, D.~A., {Hindman}, B.~W., \& {Toomre}, J. 2006,
  in ESA Special Publication, Vol. 624, Proceedings of SOHO 18/GONG 2006/HELAS
  I, Beyond the spherical Sun

\bibitem[{{Giles}(2000)}]{giles00}
{Giles}, P.~M. 2000, PhD thesis, Stanford University, USA

\bibitem[{{Gizon} \& {Birch}(2004)}]{gizon04}
{Gizon}, L., \& {Birch}, A.~C. 2004, \apj, 614, 472

\bibitem[{{Gough}(1969)}]{gough}
{Gough}, D.~O. 1969, Journal of Atmospheric Sciences, 26, 448

\bibitem[{{Hanasoge} {et~al.}(2007{\natexlab{a}}){Hanasoge}, {Duvall},
  {Derosa}, \& {Miesch}}]{hanasoge07b}
{Hanasoge}, S.~M., {Duvall}, T.~L., {Derosa}, M.~L., \& {Miesch}, M.~S.
  2007{\natexlab{a}}, in IAU Symposium, Vol. 239, IAU Symposium, ed.
  T.~{Kuroda}, H.~{Sugama}, R.~{Kanno}, \& M.~{Okamoto}, 364--369

\bibitem[{{Hanasoge} {et~al.}(2007{\natexlab{b}}){Hanasoge}, {Duvall}, \&
  {Couvidat}}]{hanasoge07a}
{Hanasoge}, S.~M., {Duvall}, Jr., T.~L., \& {Couvidat}, S. 2007{\natexlab{b}},
  \apj, 664, 1234

\bibitem[{{Hanasoge} {et~al.}(2006){Hanasoge}, {Larsen}, {Duvall}, {DeRosa},
  {Hurlburt}, {Schou}, {Roth}, {Christensen-Dalsgaard}, \& {Lele}}]{hanasoge1}
{Hanasoge}, S.~M., {Larsen}, R.~M., {Duvall}, Jr., T.~L., {DeRosa}, M.~L.,
  {Hurlburt}, N.~E., {Schou}, J., {Roth}, M., {Christensen-Dalsgaard}, J., \&
  {Lele}, S.~K. 2006, \apj, 648, 1268

\bibitem[{{Marquering} {et~al.}(1999){Marquering}, {Dahlen}, \&
  {Nolet}}]{dahlen99}
{Marquering}, H., {Dahlen}, F.~A., \& {Nolet}, G. 1999, Geophysical Journal
  International, 137, 805

\bibitem[{{Miesch} {et~al.}(2008){Miesch}, {Brun}, {De Rosa}, \&
  {Toomre}}]{miesch08}
{Miesch}, M.~S., {Brun}, A.~S., {De Rosa}, M.~L., \& {Toomre}, J. 2008, \apj,
  673, 557

\bibitem[{{Miesch} {et~al.}(2000){Miesch}, {Elliott}, {Toomre}, {Clune},
  {Glatzmaier}, \& {Gilman}}]{miesch}
{Miesch}, M.~S., {Elliott}, J.~R., {Toomre}, J., {Clune}, T.~L., {Glatzmaier},
  G.~A., \& {Gilman}, P.~A. 2000, \apj, 532, 593

\bibitem[{{Rajaguru}(2008)}]{rajaguru08}
{Rajaguru}, S.~P. 2008, ArXiv e-prints

\bibitem[{{Scherrer} {et~al.}(1995){Scherrer}, {Bogart}, {Bush}, {Hoeksema},
  {Kosovichev}, {Schou}, {Rosenberg}, {Springer}, {Tarbell}, {Title},
  {Wolfson}, {Zayer}, \& {MDI Engineering Team}}]{scherrer}
{Scherrer}, P.~H., {Bogart}, R.~S., {Bush}, R.~I., {Hoeksema}, J.~T.,
  {Kosovichev}, A.~G., {Schou}, J., {Rosenberg}, W., {Springer}, L., {Tarbell},
  T.~D., {Title}, A., {Wolfson}, C.~J., {Zayer}, I., \& {MDI Engineering Team}.
  1995, \solphys, 162, 129

\bibitem[{{Swisdak} \& {Zweibel}(1999)}]{swiss}
{Swisdak}, M., \& {Zweibel}, E. 1999, \apj, 512, 442

\bibitem[{{van Ballegooijen}(1986)}]{ball}
{van Ballegooijen}, A.~A. 1986, \apj, 304, 828

\bibitem[{{Wilson}(1987)}]{wilson}
{Wilson}, P.~R. 1987, \solphys, 110, 59

\bibitem[{{Woodard}(2007)}]{woodard}
{Woodard}, M.~F. 2007, \apj, 668, 1189

\bibitem[{{Zhao} \& {Kosovichev}(2003)}]{zhao}
{Zhao}, J., \& {Kosovichev}, A.~G. 2003, in ESA Special Publication, Vol. 517,
  GONG+ 2002. Local and Global Helioseismology: the Present and Future, ed.
  H.~{Sawaya-Lacoste}, 417--420

\end{thebibliography}

\end{document}